\pdfoutput=1
\documentclass{article} % For LaTeX2e

\newif\ifusebiblatex
\usebiblatexfalse   
\ifusebiblatex
  \makeatletter
  \expandafter\def\csname ver@natbib.sty\endcsname{2009/01/01}%
  \newcommand{\setcitestyle}[1]{}%
  \makeatother
\fi

\usepackage[preprint]{colm2026_conference}
\usepackage{lineno}
\usepackage{booktabs}
\usepackage{multirow}
\usepackage{xspace}
\usepackage{verbatim}
\usepackage{listings}
\usepackage{mdframed}
\usepackage{graphicx}
\usepackage{subcaption}
\usepackage{xcolor}
\usepackage[colorlinks=true,linkcolor=blue,citecolor=blue,urlcolor=blue]{hyperref}

\newcommand{\system}{WatchPoint\xspace}
\newcommand{\proj}[1]{\textsf{#1}}

\newmdenv[
  linewidth=0.5pt,
  linecolor=black,
  backgroundcolor=black!5,
  innertopmargin=6pt,
  innerbottommargin=6pt,
  innerleftmargin=8pt,
  innerrightmargin=8pt,
  frametitle={\textbf{Takeaway}},
  frametitleaboveskip=4pt,
]{takeaway}

\title{\system{}: Executable User Feedback for Real-World Agentic Web Development}

\author{Guanqun Yang \\
Stevens Institute of Technology \\
\texttt{guanqun.yang@outlook.com} \\
\And
Wei Yang \\
University of Texas at Dallas \\
\texttt{wei.yang@utdallas.edu} \\
\And
Xueqing Liu \\
Stevens Institute of Technology \\
\texttt{xliu127@stevens.edu} \\
}

\begin{document}

\ifcolmsubmission
\linenumbers
\fi

\maketitle
%% The preprint option stamps "Preprint. Under review." in the running head.
%% This is an archival posting, not a submission under review, so clear it.
\lhead{}

\begin{abstract}
When a professional web developer's code fails a test, they do not simply re-read the stack trace.
They open the application in a browser, click buttons, inspect computed styles, and run diagnostic commands to understand what went wrong.
Existing feedback mechanisms for coding agents rely on screenshots, LLM-as-a-judge scoring, or natural-language corrections, but few interact with the live application the way a developer would.
We introduce \system, a simulated-user system that mimics real developer behavior by generating and executing diagnostic scripts against the running application, producing structured observations that guide the coding model's retry.
Unlike prior approaches that target single-file edits or evaluate using non-executable metrics, we operate on Web-Bench, a benchmark of 50 multi-file web projects comprising 1{,}000 sequentially dependent tasks, verified by deterministic end-to-end tests.
\system recovers 57.6\% of the tasks it diagnoses, and a controlled user study confirms the simulation's realism: human testers achieve a comparable recovery rate (54.5\%), providing evidence that automated diagnostic scripts can substitute for interactive human testing on sequential web development tasks.
We further identify a pattern of capability gaps that governs when simulated-user feedback is helpful and when it should be withheld.
\end{abstract}

%%%%%%%%%%%%%%%%%%%%%%%%%%%%%%%%%%%%%%%%%%%%%%%%%%%%%%%%%%%%%%%%%%%%%%%%%%%%%%%%%%%%%%%%%%%%%%%%%%%%
% MAIN CONTENT
%%%%%%%%%%%%%%%%%%%%%%%%%%%%%%%%%%%%%%%%%%%%%%%%%%%%%%%%%%%%%%%%%%%%%%%%%%%%%%%%%%%%%%%%%%%%%%%%%%%%

\section{Introduction}
\label{sec:intro}

Coding agents increasingly operate as part of multi-agent systems in which planning, implementation, testing, and review are distributed across specialized subagents~\citep{Qian2024ChatDevCommunicativeAgents, He2025LLMBasedMultiAgentSystems, Takerngsaksiri2025HumanIntheLoopSoftwareDevelopment}.
Infrastructure for such delegation is maturing: protocols like Google's Agent-to-Agent (A2A) let agents discover and communicate~\citep{Du2025WhichLLMMultiAgent}, and frameworks like AutoGen~\citep{Wu2023AutoGenEnablingNextGen} and OpenHands~\citep{Wang2025OpenHandsOpenPlatform} provide the orchestration layer.
But a communication channel does not prescribe what to say.
\citet{Cemri2025WhyMultiAgentLLM} find that 32.3\% of multi-agent failures stem from inter-agent misalignment: agents withhold information, ignore peer input, or act on misleading feedback.
What is missing is a \emph{semantic layer}: an interface that conveys actionable diagnostic information between agents, grounded in the domain's established practices.
In web development, that practice is end-to-end testing~\citep{Xu2025WebBenchLLMCode, Zhu2025FrontendBenchBenchmarkEvaluating}: opening the application in a browser, clicking buttons, inspecting computed styles, checking server responses, and running type checkers.

\begin{comment}
Several lines of work have begun to address this feedback gap, but each covers only part of the problem.
Visual approaches supply screenshots~\citep{Lu2025WebGenAgentEnhancingInteractive, Li2025ReLookVisionGroundedRL}; conversational benchmarks measure whether agents incorporate natural-language corrections~\citep{Wu2025FronTalkBenchmarkingFrontEnd}; trajectory-level methods apply process reward models to detect and correct errors in agent traces~\citep{Gandhi2025WhenAgentsGo, Antoniades2025SWESearchEnhancingSoftware}.
However, few of these approaches interact with the live application the way a developer would: they typically evaluate single-page synthetic sites rather than multi-file real-world projects, rely on LLM-as-a-Judge scoring rather than deterministic tests~\citep{Zhang2025ArtifactsBenchBridgingVisualInteractive}, and provide one-shot screenshots with no interactive diagnosis.
\end{comment}

Several lines of work have begun to address this feedback gap, but each covers only part of the problem.
Visual approaches supply screenshots for standalone site generation~\citep{Lu2025WebGenAgentEnhancingInteractive} or train reward models for visual grounding~\citep{Li2025ReLookVisionGroundedRL}; conversational benchmarks measure whether agents incorporate natural-language corrections~\citep{Wu2025FronTalkBenchmarkingFrontEnd}; trajectory-level methods apply process reward models to detect and correct errors in agent traces~\citep{Gandhi2025WhenAgentsGo, Antoniades2025SWESearchEnhancingSoftware}.
However, few of these approaches interact with the live application the way a developer would: they typically evaluate single-page synthetic sites rather than multi-file real-world projects, rely on LLM-as-a-Judge scoring rather than deterministic tests~\citep{Zhang2025ArtifactsBenchBridgingVisualInteractive}, and provide one-shot screenshots with no interactive diagnosis.

We introduce \system, a simulated-user system that provides this semantic layer.
When a coding agent's first attempt fails the test suite, \system launches two diagnostic agents in parallel: a \emph{browser agent} that generates and executes a Playwright script to interact with the live page (clicking elements, reading computed CSS, checking DOM structure), and a \emph{terminal agent} that generates and executes a bash script to inspect server logs, check file contents, and run type checkers.
Both scripts produce structured key-value observations that are fed back to the coding agent for a targeted retry (\S\ref{sec:method}).
The design mirrors the feedback loop a human tester would provide: not merely ``the page looks wrong,'' but ``\texttt{flex-wrap} is \texttt{wrap}, expected \texttt{wrap-reverse}.''

We evaluate \system{} on Web-Bench~\citep{Xu2025WebBenchLLMCode}, a benchmark of 50 real-world web projects with 1{,}000 sequential tasks that remains challenging for frontier agents---Claude Code, for instance, achieves only 13.4\% Pass@1.
Our primary experiments use the OpenHands scaffold~\citep{Wang2025OpenHandsOpenPlatform} with two coding models (MiniMax M2.7 and GLM-5) and two diagnostic models (GLM-5 and GPT-5.4) to study how feedback quality interacts with the coding model's capability.

Our experiments yield three findings:
\begin{enumerate}
% Old: \item \textbf{Recovery rate ...} ... gaining up to 28\% on Pass@2.
\item \textbf{Recovery rate (\S\ref{sec:experiment}, Table~\ref{tab:category-results}).} \system recovers 57.6\% of the tasks it diagnoses, with state-management projects gaining 7.5\% on Pass@2.
\item \textbf{Simulation realism (\S\ref{sec:experiment}, Table~\ref{tab:user-study-results}).} A user study shows that this recovery rate is comparable to human feedback quality (54.5\%); on the four projects where both humans and \system can inspect the running application, \system matches the stronger human participant.
% Old: \item \textbf{Capability gap ...} ... diagnostic feedback can introduce noise.
\item \textbf{Capability gap (\S\ref{sec:experiment}, Table~\ref{tab:main-results}).} Feedback effectiveness depends on the gap between the coding model and the diagnostic agent: when the coding model is weak, \system provides clear value, but when it is already strong enough to self-correct from test errors alone, diagnostic feedback can introduce noise.
\end{enumerate}
% Old: These results carry practical implications ... staying out of the way of competent ones.
These results carry practical implications for hierarchical agent systems: the semantic layer must be quality-aware, routing diagnostics to struggling subagents while staying out of the way of competent ones.

\section{Related Work}
\label{sec:related}

\subsection{Agentic Coding Benchmarks and Workflows}
SWE-bench~\citep{Jimenez2023SWEbenchCanLanguage} introduces the task of resolving real-world GitHub issues against executable test suites, spawning numerous variants and extensions~\citep{OpenAI2024IntroducingSWEbenchVerified, Deng2025SWEBenchProCan, Wang2025SWEBenchFrameworkScalable}.
Web-Bench~\citep{Xu2025WebBenchLLMCode} extends this paradigm to full-stack web development by organizing 50 projects into sequences of 20 ordered tasks verified by expert-curated Playwright end-to-end tests, where early failures cascade through the project.
On the agent side, SWE-agent~\citep{Yang2024SWEagentAgentComputerInterfaces} proposes the Agent--Computer Interface (ACI) that equips language models with file-navigation and editing commands, while Agentless~\citep{Xia2024Agentless} demonstrates that structured ``localize-then-repair'' prompting can match autonomous agents without iterative tool use.
%%%%%%%%%%%%%%%%%%%%%%%%%%%%%%%%%%%%%%%%%%%%%%%%%%%%%%%%%%%%%%%%%%%%%%%
% BEGIN NOT-IN-OVERLEAF: InspectCoder citation + three-way parallel
%%%%%%%%%%%%%%%%%%%%%%%%%%%%%%%%%%%%%%%%%%%%%%%%%%%%%%%%%%%%%%%%%%%%%%%
InspectCoder~\citep{Wang2025InspectCoderDynamicAnalysisEnabled} uses debugger breakpoints and variable inspection for LLM self-repair, providing executable diagnosis at the code level rather than the application level.
Our work draws on these lines: like Agentless, our diagnostic agents use structured, single-shot generation rather than iterative exploration; like InspectCoder, we use executable observations rather than LLM judgment; and like multi-agent systems, we decompose the feedback loop into specialized browser and terminal phases.
%%%%%%%%%%%%%%%%%%%%%%%%%%%%%%%%%%%%%%%%%%%%%%%%%%%%%%%%%%%%%%%%%%%%%%%
% END NOT-IN-OVERLEAF
%%%%%%%%%%%%%%%%%%%%%%%%%%%%%%%%%%%%%%%%%%%%%%%%%%%%%%%%%%%%%%%%%%%%%%%

\subsection{User Simulation and Feedback}

% Old: \citet{Gandhi2025WhenAgentsGo} show that a process reward model must be stronger than the policy it supervises, a capability-gap constraint that our cross-model experiments confirm for interactive diagnostic feedback.
\citet{Gandhi2025WhenAgentsGo} show that a process reward model must be stronger than the policy it supervises, a capability-gap constraint that our cross-model experiments confirm for interactive diagnostic feedback.
%%%%%%%%%%%%%%%%%%%%%%%%%%%%%%%%%%%%%%%%%%%%%%%%%%%%%%%%%%%%%%%%%%%%%%%
% BEGIN NOT-IN-OVERLEAF: SWE-PRM learned-vs-generated distinction
%%%%%%%%%%%%%%%%%%%%%%%%%%%%%%%%%%%%%%%%%%%%%%%%%%%%%%%%%%%%%%%%%%%%%%%
SWE-PRM~\citep{Gandhi2025WhenAgentsGo} applies a frozen LLM as a learned process reward over agent trajectories; \system instead generates task-specific diagnostic scripts at inference time, differing in whether the feedback signal is learned or generated on the fly.
%%%%%%%%%%%%%%%%%%%%%%%%%%%%%%%%%%%%%%%%%%%%%%%%%%%%%%%%%%%%%%%%%%%%%%%
% END NOT-IN-OVERLEAF
%%%%%%%%%%%%%%%%%%%%%%%%%%%%%%%%%%%%%%%%%%%%%%%%%%%%%%%%%%%%%%%%%%%%%%%
WebGen-Agent~\citep{Lu2025WebGenAgentEnhancingInteractive} iteratively refines websites using VLM screenshot scoring and interactive GUI-agent testing, but targets standalone sites generated from scratch rather than modifications to existing codebases with sequential task dependencies.
\citet{Zhang2025BootstrappingVisualAssistant} train visual assistants entirely from synthetic user interactions, and HAICOSYSTEM~\citep{Zhou2025HAICOSYSTEMEcosystemSandboxing} demonstrates that sandboxed, multi-turn interactions surface up to $3\times$ more safety risks than single-turn probes.
\system builds on these insights by coupling executable diagnostic scripts with a simulated-user loop designed for multi-step, interactive web applications, bridging the gap between static visual scoring and dynamic software development.

\section{\system System}\label{sec:method}

\begin{figure*}[t]
  \centering
  \includegraphics[width=\textwidth]{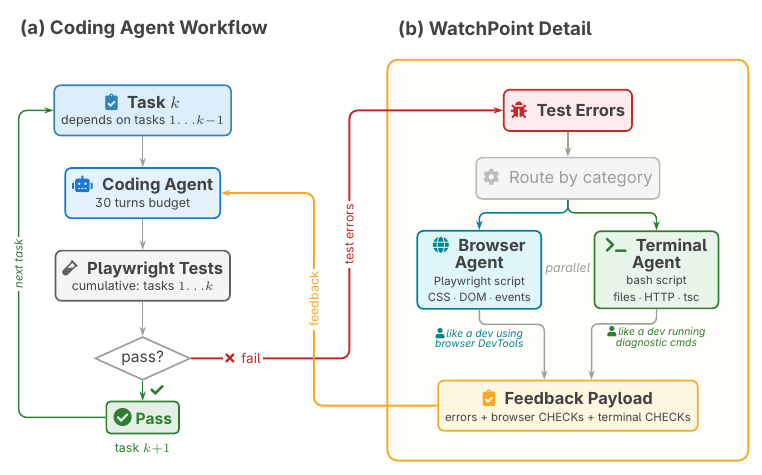}
  \caption{Overview of the \system workflow. (a)~The coding agent receives a task, iteratively edits code, and runs Playwright E2E tests. On failure, \system generates diagnostic feedback via two parallel agents. (b)~Detail of the diagnostic mechanism: a browser agent executes a Playwright script to inspect the live page, a terminal agent runs a bash script to check workspace state, and the combined structured observations are fed back to the coding model for a retry attempt.}
  \label{fig:system}
\end{figure*}

\subsection{Agentic Coding Runtime Environment}
\label{subsec:runtime}

Evaluating simulated-user feedback requires a benchmark with sequential, multi-file tasks in which early failures cascade.
Single-function benchmarks like HumanEval~\citep{Chen2021EvaluatingLargeLanguage} and MBPP~\citep{Austin2021ProgramSynthesisLarge} lack this property, as do isolated-issue benchmarks like SWE-bench~\citep{Jimenez2023SWEbenchCanLanguage}.
We adopt Web-Bench~\citep{Xu2025WebBenchLLMCode}, which provides 50 real-world web projects spanning frontend frameworks (React, Vue, Angular, Svelte), state management (Redux, Zustand, Jotai), CSS tooling (Tailwind, Styled-Components), build systems (Vite, Webpack), full-stack frameworks (Next.js, Express.js, Fastify), and database ORMs (Sequelize, Prisma, Lowdb).
Each project defines a sequence of 20 development tasks of increasing complexity, totaling 1{,}000 tasks.
Tasks are verified by deterministic Playwright end-to-end tests that exercise the running application, not by LLM-as-a-judge scoring.
A strict sequential dependency chain governs the evaluation: if the agent fails task $k$ after two attempts, all subsequent tasks $k{+}1$ through $20$ are skipped.

We report two complementary metrics following \cite{Xu2025WebBenchLLMCode}.
$$\mathrm{Pass@1} = \frac{n_{\mathrm{pass},1}}{N}, \qquad \mathrm{Pass@2} = \frac{n_{\mathrm{pass},2}}{N},$$
where $n_{\mathrm{pass},1}$ is the number of tasks passing all E2E tests on the first attempt, $n_{\mathrm{pass},2}$ is the total number of tasks passing within two attempts (including retries), and $N$ is the total number of tasks.
Pass@1 reflects unassisted coding ability; the gap between Pass@2 and Pass@1 isolates the differential value of the retry and its associated feedback.
%%%%%%%%%%%%%%%%%%%%%%%%%%%%%%%%%%%%%%%%%%%%%%%%%%%%%%%%%%%%%%%%%%%%%%%
% BEGIN NOT-IN-OVERLEAF: Recovery rate definition (moved here from §4)
%%%%%%%%%%%%%%%%%%%%%%%%%%%%%%%%%%%%%%%%%%%%%%%%%%%%%%%%%%%%%%%%%%%%%%%
We also report the \textbf{per-task recovery rate}: the fraction of tasks that failed attempt~1 and were retried (not skipped by the sequential stopping rule) that passed on attempt~2.
This metric is not subject to cascade amplification and directly measures feedback quality.
%%%%%%%%%%%%%%%%%%%%%%%%%%%%%%%%%%%%%%%%%%%%%%%%%%%%%%%%%%%%%%%%%%%%%%%
% END NOT-IN-OVERLEAF
%%%%%%%%%%%%%%%%%%%%%%%%%%%%%%%%%%%%%%%%%%%%%%%%%%%%%%%%%%%%%%%%%%%%%%%

\subsection{Agentic Coding Workflow}
\label{subsec:agent-workflow}

Our primary agent scaffold is the OpenHands SDK (v1)~\citep{Wang2025OpenHandsOpenPlatform}.
The agent has access to two tools: a \textsc{Terminal} for executing bash commands in a Docker-sandboxed environment, and a \textsc{FileEditor} for viewing, creating, and performing string-replacement edits on source files.
Each attempt is budgeted at 30 iterations, where one iteration corresponds to one LLM call.
To enable parallel evaluation, we re-engineered the original Web-Bench harness into a Docker-based setup with per-project filesystem and port isolation, completing each configuration in 2--4 hours with 10 parallel workers (Appendix~\ref{sec:infra}).

The original Web-Bench framework retries by passing raw test errors~\citep{Xu2025WebBenchLLMCode}.
We improve this with a \emph{warm-start} strategy: the second attempt continues the existing conversation, appending test errors and any diagnostic observations to the context window, preserving the agent's prior reasoning.
The retry feedback consists of (1)~filtered test errors from the Playwright runner (Appendix~\ref{sec:feedback-payload}), (2)~structured browser observations from \system (\S\ref{subsec:simulated-user}), and (3)~structured terminal observations from \system (\S\ref{subsec:simulated-user}).
When \system is disabled, only test errors are included.
Prompt templates are in Appendix~\ref{sec:prompts}.

\subsection{Simulated User (\system)}
\label{subsec:simulated-user}

The simulated user is the core contribution of this work.
When the coding agent fails a task on its first attempt, \system launches two lightweight diagnostic agents in parallel: a \emph{browser agent} and a \emph{terminal agent}.
The diagnostic agents are independent of the coding model and can be powered by a different LLM, enabling us to study cross-model interactions.

\paragraph{Browser agent.}
Given the failing test assertions, an LLM generates a Playwright script that launches headless Chromium and checks what the page \emph{actually} renders.
For each failed assertion, the script navigates to the relevant URL, interacts with page elements as needed (e.g., clicking buttons, filling forms, hovering over targets), and inspects concrete properties (e.g., computed CSS values, DOM structure, text content, element visibility, and event-handler behavior).
Findings are logged as structured JSON:

\begin{lstlisting}
console.log(JSON.stringify({
  n: "check_name", v: "value"
}));
\end{lstlisting}

\noindent Each observation maps directly to a specific test assertion, providing the coding model with a precise diagnosis rather than a vague ``the page looks wrong'' signal~\citep{Dai2026FeedbackEvalBenchmarkEvaluating}.

\paragraph{Terminal agent.}
A second LLM generates a bash script that diagnoses server-side and build-system failures.
The script inspects file contents relevant to failing tests, starts backend servers and issues HTTP requests for full-stack projects, and runs type-checking passes (e.g., \texttt{npx tsc --noEmit}) for TypeScript projects.
Results are logged in the same JSON format, allowing both agents' outputs to be combined into a single retry message.

\paragraph{Why two agents.}
Frontend failures (incorrect CSS, missing DOM elements, broken event handlers) are observable only when rendering the page in a browser; backend and build failures (missing dependencies, type errors, incorrect API responses) are observable only via the filesystem and running processes.
A browser agent alone cannot detect a missing npm dependency; a terminal agent alone cannot detect a misaligned CSS layout. Together, the two agents cover both frontend and backend failure modes.

\paragraph{Why single-shot.}
Each diagnostic agent makes a single LLM call to generate a script, which is then run once against the live application; the raw output (e.g., \texttt{flex-wrap: wrap}) is passed directly to the coding model without intermediate LLM interpretation.
An iterative loop would introduce additional LLM calls to observe, decide, and synthesize, each of which could be a point at which the model hallucinates.
\citet{Seshadri2026LostSimulationLLMSimulated} confirm this risk: LLM-simulated users accumulate systematic biases across turns, with an expected calibration error of 15.1 between simulated and real user assessments.
Our design avoids this drift---the LLM decides \emph{what to measure} but never \emph{interprets what was measured}.
The coding model sees ``\texttt{flex-wrap: wrap}'' (a runtime fact), not ``the layout appears incorrect'' (an LLM judgment).
%%%%%%%%%%%%%%%%%%%%%%%%%%%%%%%%%%%%%%%%%%%%%%%%%%%%%%%%%%%%%%%%%%%%%%%
% BEGIN NOT-IN-OVERLEAF: Expanded v1 hallucination (was 1 sentence)
%%%%%%%%%%%%%%%%%%%%%%%%%%%%%%%%%%%%%%%%%%%%%%%%%%%%%%%%%%%%%%%%%%%%%%%
% Old: Our early experiments validate this: a two-phase design with LLM interpretation introduced hallucinations on \proj{calculator} (Appendix~\ref{sec:v1-hallucination}); removing interpretation eliminated the failure.
Our early experiments confirm this risk.
An earlier two-phase design collected browser observations with a hardcoded Playwright script and passed them to GLM-5 for interpretation.
On the \proj{calculator} project, GLM-5 fabricated a ``Blog'' page title and phantom server errors, reducing Pass@1 from 11/20 to 4/20 across 6 of 7 diagnostics-triggered tasks (Appendix~\ref{sec:v1-hallucination}).
Removing the interpretation step eliminated the failure.
%%%%%%%%%%%%%%%%%%%%%%%%%%%%%%%%%%%%%%%%%%%%%%%%%%%%%%%%%%%%%%%%%%%%%%%
% END NOT-IN-OVERLEAF
%%%%%%%%%%%%%%%%%%%%%%%%%%%%%%%%%%%%%%%%%%%%%%%%%%%%%%%%%%%%%%%%%%%%%%%
%%%%%%%%%%%%%%%%%%%%%%%%%%%%%%%%%%%%%%%%%%%%%%%%%%%%%%%%%%%%%%%%%%%%%%%
% BEGIN NOT-IN-OVERLEAF: P4.1 + P4.2 script success rate and cost
%%%%%%%%%%%%%%%%%%%%%%%%%%%%%%%%%%%%%%%%%%%%%%%%%%%%%%%%%%%%%%%%%%%%%%%
The single-shot architecture also keeps overhead low: both scripts run as external subprocesses in parallel (${\sim}$15\,s wall-clock), consume zero coding-model iterations, and add less than 2\% to the coding model's API cost (Appendix~\ref{sec:diagnostic-overhead}).
In practice, 59 of 66 diagnostic runs (89\%) produced at least one valid CHECK observation; the 7 failures occurred on projects where the dev server did not expose a browser-inspectable page.
%%%%%%%%%%%%%%%%%%%%%%%%%%%%%%%%%%%%%%%%%%%%%%%%%%%%%%%%%%%%%%%%%%%%%%%
% END NOT-IN-OVERLEAF
%%%%%%%%%%%%%%%%%%%%%%%%%%%%%%%%%%%%%%%%%%%%%%%%%%%%%%%%%%%%%%%%%%%%%%%

\paragraph{Selective activation.}
Not all project categories benefit equally from browser inspection.
\system classifies each project by parsing its \texttt{package.json} dependencies (e.g., the presence of \texttt{react} indicates a UI project, \texttt{prisma} indicates a database project), following the same convention used by deployment platforms such as Vercel\footnote{\url{https://vercel.com/docs/builds/configure-a-build\#framework-preset}} and Netlify\footnote{\url{https://docs.netlify.com/build/configure-builds/manage-dependencies/}}.
Diagnostic agents are enabled only for Standards, UI, and State projects (92\% gating accuracy), where DOM and CSS inspection is most informative.
The full classification rules are in Appendix~\ref{sec:classification}.

\subsection{Contrast with Prior Feedback Approaches}
\label{subsec:method-contrast}

Prior feedback mechanisms provide visual observations scored by LLM critics~\citep{Lu2025WebGenAgentEnhancingInteractive, Li2025ReLookVisionGroundedRL}, conversational comments on rendered output~\citep{Wu2025FronTalkBenchmarkingFrontEnd}, or trajectory-level process rewards~\citep{Gandhi2025WhenAgentsGo}.
\system instead performs \emph{active diagnosis}: it opens the live application, clicks elements, reads computed styles, curls API endpoints, and measures DOM properties against expected values, mirroring a professional tester who systematically probes behavior rather than glancing at a page.

\section{Experiments}\label{sec:experiment}

\subsection{Research Questions}
Our experiments investigate the performance of agentic coding systems and the impact of user simulation. 
Specifically, we address three research questions:
\begin{description}
    \item[\textbf{RQ1.}] How do agentic coding systems perform on sequential web development tasks?
    \item[\textbf{RQ2.}] How effective is the simulated user at recovering failed tasks?
    \item[\textbf{RQ3.}] How does the simulated user compare to real human feedback?
\end{description}

\subsection{Experiment Setup}

\paragraph{Benchmark.}
As described in \S\ref{subsec:runtime}, we evaluate on Web-Bench~\citep{Xu2025WebBenchLLMCode}, which comprises 50 real-world open-source web projects with 20 sequential tasks each (1{,}000 tasks total).
Tasks are ordered by dependency: task $k$ assumes the correct completion of tasks $1, \ldots, k{-}1$, so early failures cascade to all subsequent tasks.

\paragraph{Configurations.}
Table~\ref{tab:main-results} summarizes the six configurations we evaluated, varying the coding model, the simulated user model, and the agent scaffold.
Configuration C0 uses the Claude Code harness with Sonnet~4.6 as the coding model and serves as a difficulty baseline for the benchmark; no simulated user is applied.
Configurations C1--C5 use the OpenHands scaffold~\citep{Wang2025OpenHandsOpenPlatform}, which provides a Docker-sandboxed environment with terminal access, file editing, and browser interaction.
C2, C3, and C5 add a \system simulated user of varying capability.
For all configurations, we set the temperature to $0.0$ and disabled extended reasoning to ensure deterministic, cost-controlled generation, with a budget of 30 iterations per task.
Complete model hyperparameters are listed in Appendix~\ref{sec:hyperparams}, Table~\ref{tab:hyperparams}.

\paragraph{Models.}
We selected coding models that are Pareto-optimal on the coding-capability vs.\ cost frontier at the time of the experiment: no cheaper model achieves higher coding performance at any tier.\footnote{\url{https://artificialanalysis.ai/models/capabilities/coding}}
MiniMax M2.7 represents the mid-tier, GLM-5 the high-tier, and GPT-5.4 the top-tier (used exclusively as a simulated user).
Claude Opus~4.6 was considered but proved prohibitively slow due to rate limits under the highest subscription tier (\$200/month), making it infeasible for 1{,}000-task experiments.

\paragraph{Infrastructure.}
Each configuration completes in 2--4 hours across 10 parallel Docker workers with a 90-minute per-project timeout (Appendix~\ref{sec:infra}).
We report P@$k$ as the fraction of tasks where at least one of $k$ attempts passes all tests; P@1 measures first-attempt success and P@2 measures success after at most one retry.
Further details are in Appendix~\ref{sec:hyperparams}.

\subsection{RQ1: Agentic Performance on Web-Bench}

\begin{table}[t]
\centering
\caption{Aggregate results on Web-Bench. P@1 is the first-attempt pass rate; P@2 is the pass rate after at most one retry. Best agentic P@1 and P@2 are \textbf{bolded}.}
\label{tab:main-results}
\small
\begin{tabular}{clllrr}
\toprule
\textbf{ID} & \textbf{Coding Model} & \textbf{Simulated User} & \textbf{Scaffold} & \textbf{P@1} & \textbf{P@2} \\
\midrule
--- & \citet{Xu2025WebBenchLLMCode}&---& non-agentic$^\dagger$ & 25.1\% & 35.3\% \\
\midrule
C0 & Sonnet 4.6 & --- & Claude Code & 13.4\% & 21.5\% \\
\midrule
C1 & MiniMax M2.7 & --- & OpenHands & 18.0\% & 26.6\% \\
C2 & MiniMax M2.7 & GLM-5 & OpenHands & 16.4\% & 28.1\% \\
C3 & MiniMax M2.7 & GPT-5.4 & OpenHands & 15.6\% & 27.9\% \\
C4 & GLM-5 & --- & OpenHands & \textbf{25.0\%} & \textbf{44.5\%} \\
C5 & GLM-5 & GPT-5.4 & OpenHands & 22.0\% & 37.7\% \\
\bottomrule
\multicolumn{6}{l}{\scriptsize $^\dagger$ Best-of-5 sampling with temp.\ $\geq 0.4$; our rows use temp.\ $0$.}
\end{tabular}
\end{table}

Table~\ref{tab:main-results} presents the aggregate results across all six configurations alongside the non-agentic baseline reported by \citet{Xu2025WebBenchLLMCode}.

\paragraph{Agentic workflows rival heavy sampling.}
The \citet{Xu2025WebBenchLLMCode} baseline uses best-of-5 stochastic runs (temperature 0.4--1.0); our configurations use temperature 0 for reproducibility, though some variance remains from agentic non-determinism and cascade sensitivity (Appendix~\ref{sec:variance}).
%%%%%%%%%%%%%%%%%%%%%%%%%%%%%%%%%%%%%%%%%%%%%%%%%%%%%%%%%%%%%%%%%%%%%%%
% BEGIN NOT-IN-OVERLEAF: "matches the baseline" fix + C0 scaffold footnote
%%%%%%%%%%%%%%%%%%%%%%%%%%%%%%%%%%%%%%%%%%%%%%%%%%%%%%%%%%%%%%%%%%%%%%%
Despite this, GLM-5 matches the baseline at 25.0\% P@1 and substantially exceeds it on P@2 (44.5\% vs.\ 35.3\%), showing that agentic tool use with warm-start retries can outperform non-agentic best-of-5 sampling.
MiniMax M2.7 (18.0\% P@1) and Sonnet~4.6 (13.4\% P@1) fall behind, reflecting weaker capabilities compounded by the cascade structure.\footnote{C0's lower result also reflects scaffold differences: Claude Code uses cold-start retries (fresh conversation per attempt) while OpenHands uses warm-start (continuing the existing conversation), and the two scaffolds expose different tool sets.}
%%%%%%%%%%%%%%%%%%%%%%%%%%%%%%%%%%%%%%%%%%%%%%%%%%%%%%%%%%%%%%%%%%%%%%%
% END NOT-IN-OVERLEAF
%%%%%%%%%%%%%%%%%%%%%%%%%%%%%%%%%%%%%%%%%%%%%%%%%%%%%%%%%%%%%%%%%%%%%%%

\paragraph{Self-correction and Web-Bench difficulty.} GLM-5 achieves 44.5\% P@2 without any simulated-user feedback, recovering a substantial fraction of tasks from test errors alone; any simulated-user intervention must add value beyond this strong self-correction baseline.
The modest Sonnet~4.6 result (13.4\% P@1) confirms that Web-Bench remains challenging even for frontier models, with early errors compounding rapidly through the cascade.

\begin{takeaway}
Even frontier agentic systems find Web-Bench challenging. The cascade structure penalizes compounding errors, and no agentic configuration substantially outperforms the non-agentic baseline on first-attempt pass rate P@1.
\end{takeaway}

\subsection{RQ2: Simulated User Effectiveness}

\begin{table*}[t]
\centering
\caption{Per-category results for M2.7 baseline (C1) vs.\ M2.7 + GLM-5 simulated user (C2).
``ON/OFF'' = simulated user enabled/disabled.
P@2 columns show aggregate pass rates.
Recovery columns show the per-task retry success: ``Retried'' = tasks that failed attempt~1 and were retried (not skipped); ``Rec.'' = passed on attempt~2; ``Rate'' = Rec./Retried.}
\label{tab:category-results}
\label{tab:recovery}
\small
\begin{tabular}{lccrrrrrr}
\toprule
& & & \multicolumn{3}{c}{\textbf{P@2 (\% of tasks)}} & \multicolumn{3}{c}{\textbf{C2 Recovery (tasks)}} \\
\cmidrule(lr){4-6} \cmidrule(lr){7-9}
\textbf{Category} & \textbf{\#Projects} & \textbf{\system} & \textbf{C1} & \textbf{C2} & \textbf{$\Delta$} & \textbf{Retried} & \textbf{Rec.} & \textbf{Rate} \\
\midrule
Standards & 18 & ON & 23.3\% & 24.4\% & +1.1 & 44 & 26 & 59.1\% \\
UI & 6 & ON & 24.2\% & 28.3\% & +4.2 & 15 & 9 & 60.0\% \\
State & 4 & ON & 11.2\% & 18.8\% & +7.5 & 7 & 3 & 42.9\% \\
\midrule
\emph{ON total} & 28 & & \emph{21.8\%} & \emph{24.5\%} & \emph{+2.7} & \emph{66} & \emph{38} & \emph{57.6\%} \\
\midrule
CSS & 6 & OFF & 30.0\% & 30.8\% & +0.8 & 12 & 6 & 50.0\% \\
Build & 2 & OFF & 20.0\% & 25.0\% & +5.0 & 3 & 1 & 33.3\% \\
Fullstack & 5 & OFF & 38.0\% & 17.0\% & $-$21.0 & 7 & 2 & 28.6\% \\
Database & 4 & OFF & 36.2\% & 31.2\% & $-$5.0 & 7 & 3 & 42.9\% \\
Other & 3 & OFF & 45.0\% & 60.0\% & +15.0 & 8 & 5 & 62.5\% \\
\midrule
\emph{OFF total} & 20 & & \emph{34.5\%} & \emph{31.2\%} & \emph{$-$3.2} & \emph{37} & \emph{17} & \emph{45.9\%} \\
\bottomrule
\end{tabular}
\end{table*}

\paragraph{Recovery rate.}
The most direct measure of feedback quality is the \textbf{per-task recovery rate}: among the tasks that failed attempt~1 and were retried, how many passed on attempt~2?
Table~\ref{tab:category-results} shows that on simulated-user-enabled categories (ON), C2 recovers 38 of 66 retried tasks (\textbf{57.6\%}), compared to 45.1\% for C1 on the same categories using only test errors.
On disabled categories (OFF), the simulated user never runs, so any difference between C1 and C2 is pure noise.
The per-task recovery rate confirms this: C2 recovers 45.9\% of retried tasks, nearly identical to C1's 46.2\%.
The aggregate P@2, however, drops by 3.2pp.
The reason is the sequential stopping rule: in C2's run, the Fullstack category happened to fail earlier tasks on attempt~1 ($-$21pp), which skipped all downstream tasks and dragged down the total.
The simulated user played no role in these projects.

\begin{comment}
\paragraph{Per-category gains.}
On simulated-user-enabled categories (Standards, UI, State), P@2 improved from 21.8\% to 24.5\% (+2.7pp, +15 tasks).
State-management projects see the largest relative gain (+7.5pp), followed by UI (+4.2pp) and Standards (+1.1pp).
A full per-project breakdown is provided in Appendix~\ref{sec:per-project-results}.

The strongest individual project gains (Appendix~\ref{sec:per-project-results}) are \proj{mobx} (+40pp), \proj{zustand} (+35pp), \proj{expression-editor} (+35pp), and \proj{jotai} (+30pp).
Because Web-Bench tasks are sequential, recovering a single early task can unlock many downstream tasks: in \proj{esmodule}, the simulated user recovered task~1, which unblocked 8 further tasks (0/20 $\to$ 9/20).
\end{comment}

\paragraph{Per-category gains.}
On simulated-user-enabled categories (Standards, UI, State), P@2 improved from 21.8\% to 24.5\% (+2.7pp, +15 tasks).
State-management projects see the largest relative gain (+7.5pp), followed by UI (+4.2pp) and Standards (+1.1pp).
A full per-project breakdown is provided in Appendix~\ref{sec:per-project-results}.

%%%%%%%%%%%%%%%%%%%%%%%%%%%%%%%%%%%%%%%%%%%%%%%%%%%%%%%%%%%%%%%%%%%%%%%
% BEGIN NOT-IN-OVERLEAF: Corrected project gains (was mobx/jotai/expr-editor)
%%%%%%%%%%%%%%%%%%%%%%%%%%%%%%%%%%%%%%%%%%%%%%%%%%%%%%%%%%%%%%%%%%%%%%%
The strongest individual project gains (Appendix~\ref{sec:per-project-results}) are \proj{esmodule} (+45pp), \proj{react} (+35pp), \proj{zustand} (+30pp), and \proj{dom} (+25pp).
Because Web-Bench tasks are sequential, recovering a single early task can unlock many downstream tasks: in \proj{esmodule}, the simulated user recovered an early task, which unblocked 8 further tasks (1/20 $\to$ 9/20).
%%%%%%%%%%%%%%%%%%%%%%%%%%%%%%%%%%%%%%%%%%%%%%%%%%%%%%%%%%%%%%%%%%%%%%%
% END NOT-IN-OVERLEAF
%%%%%%%%%%%%%%%%%%%%%%%%%%%%%%%%%%%%%%%%%%%%%%%%%%%%%%%%%%%%%%%%%%%%%%%

\paragraph{Aggregate vs.\ per-task metrics.}
Despite the strong per-task recovery rate, the aggregate P@2 improvement is modest: +1.5pp with the GLM-5 simulated user (26.6\% $\to$ 28.1\%) and +1.3pp with GPT-5.4 (26.6\% $\to$ 27.9\%).
This discrepancy arises because the sequential metric amplifies early-task variance: a single failure at task~3 can cancel out 17 downstream successes, leading to a per-project standard deviation of approximately 24pp (Appendix~\ref{sec:variance}).
Several projects show large drops (e.g., \proj{draw} 15$\to$0), but log inspection reveals these trace to attempt-1 variance: the coding model generated different code on its first attempt across runs, and the sequential stopping rule amplified the difference (Appendix~\ref{sec:log-excerpts}).
Projects like \proj{expressjs} and \proj{sequelize} that appear in the hurt column received no simulated user feedback; their drops are pure run-to-run noise.

\paragraph{A stronger diagnostic model does not help more.}
% Old: A surprising finding is that GPT-5.4 (C3) achieves the same aggregate P@2 ...
A surprising finding is that GPT-5.4 (C3) achieves similar aggregate P@2 to GLM-5 (C2), despite being substantially more capable.
The two diagnostic models help on different projects: GLM-5 excels on Standards projects with DOM/CSS issues, while GPT-5.4 excels on State and SVG projects.
Neither advantage is sufficiently consistent to dominate the other, and the cascade metric cancels out project-level differences.
This suggests that diagnostic quality depends on the type of failure, not on the general coding capability of the diagnostic model (Appendix~\ref{sec:judge-comparison}).

\paragraph{Deployment conditions.}
For MiniMax M2.7 (the weaker coding model), both simulated users improve P@2, and the 57.6\% recovery rate demonstrates clear value---M2.7 frequently produces obvious failures (missing imports, malformed CSS) that a stronger diagnostic agent reliably identifies.
For GLM-5 (the stronger coding model), the picture reverses: adding a GPT-5.4 simulated user \emph{decreases} P@2 from 44.5\% to 37.7\% (C4$\to$C5).
Log inspection traces this regression to two failure modes: (1)~diagnostic scripts that target the wrong port or URL, producing irrelevant CHECKs that distract the coding model (e.g., the \proj{draw} collapse in Appendix~\ref{sec:log-excerpts}), and (2)~CHECKs that correctly identify surface symptoms but misattribute the root cause, leading the model to modify working code.
The pattern echoes \citet{Gandhi2025WhenAgentsGo}, who show that a process reward model must be stronger than the policy it monitors: when the coding model is already strong enough to self-correct from test errors alone, additional diagnostic feedback introduces more noise than signal.
We note that the 6.8pp drop overlaps with run-to-run variance (Appendix~\ref{sec:variance}); confirming this pattern with additional model pairs is an important direction.

\subsection{RQ3: Comparison with Human Feedback}

\begin{table*}[tp]
\centering
\caption{User study overview. (a)~Project selection with automated P@2 for reference. (b)~Human vs.\ \system{} comparison. All numbers are tasks out of 20 per project.}
\label{tab:user-study}
\small
\begin{minipage}[t]{0.47\textwidth}
\centering
\subcaption{Project selection.}
\label{tab:user-study-projects}
\footnotesize
\begin{tabular}{lrrr}
\toprule
\textbf{Project} & \textbf{C1 P@1} & \textbf{C1 P@2} & \textbf{C2 P@2} \\
\midrule
calculator & 10 & 10 & 12\,{\tiny(+2)} \\
flex & 3 & 8 & 10\,{\tiny(+2)} \\
expr-editor & 7 & 7 & 14\,{\tiny(+7)} \\
form & 1 & 1 & 5\,{\tiny(+4)} \\
chart & 0 & 6 & 6\,{\tiny(+0)} \\
\midrule
\emph{Total (/100)} & 21 & 32 & 47\,{\tiny(+15)} \\
\bottomrule
\end{tabular}
\end{minipage}%
\hspace{0.03\textwidth}%
\begin{minipage}[t]{0.47\textwidth}
\centering
\subcaption{Human vs.\ \system{} (P@2).}
\label{tab:user-study-results}
\footnotesize
\setlength{\tabcolsep}{3pt}
\begin{tabular}{lrrrr}
\toprule
\textbf{Project} & \textbf{C1} & \textbf{C2} & \textbf{Hum.\ A} & \textbf{Hum.\ B} \\
\midrule
calculator & 10 & 12 & 11 & 5 \\
chart & 6 & 6 & 4 & 7 \\
flex & 8 & 10 & 8 & 12 \\
form & 1 & 5 & 5 & 9 \\
\midrule
\emph{4-project\ (/80)} & 25 & 33 & 28 & 33 \\
\midrule
expr-editor$^\dagger$ & 7 & 14 & 0 & 0 \\
\bottomrule
\end{tabular}
\\[4pt]
{\scriptsize $^\dagger$ Dev server unreachable via SSH; excluded from subtotal.}
\end{minipage}
\end{table*}

The 57.6\% recovery rate shows that automated diagnostics are effective; we now ask whether this is comparable to real human feedback.
Two graduate students replace the simulated user in the M2.7 + OpenHands workflow on five pure HTML/CSS/JavaScript projects (\proj{calculator}, \proj{flex}, \proj{expression-editor}, \proj{form}, \proj{chart}) that render directly in a browser without a framework build step (Table~\ref{tab:user-study-projects}).

\paragraph{Protocol.}
Each participant runs configuration C1 with one change: on first-attempt failure, the participant inspects the application in a browser (via SSH tunnel) and provides free-form feedback (e.g., ``the sqrt button appears but clicking it does nothing''), which replaces the structured CHECK observations in the retry prompt.
All other parameters are identical to C1. Full protocol details are in Appendix~\ref{sec:user-study-protocol}.

\paragraph{Results.}
Table~\ref{tab:user-study-results} compares the two participants against the baseline (C1) and \system (C2).\footnote{Of the five planned projects, four were successfully evaluated by both participants. The \proj{expression-editor} dev server was unreachable through the participants' SSH tunnels due to VPN-related port-forwarding restrictions on the university network. We report \proj{expression-editor} separately and exclude it from the 4-project comparison.}
On the four projects where both humans and \system can inspect the running application, \system and Human~B achieve identical P@2 (33/80), while Human~A reaches 28/80.
The most direct measure of feedback quality is the \textbf{per-task recovery rate}: given that feedback is provided, what fraction of failing tasks does the coding model fix?
\system recovers 57.6\% of the tasks it diagnoses; the two humans recover 44\% and 62\% respectively (average 54.5\%).
% Old: This comparable rate establishes that \system produces feedback of similar quality to real human interaction, providing concrete evidence for simulation realism.
This comparable rate provides preliminary evidence that \system produces feedback of similar quality to interactive human testing, though we note that the study is small: two graduate students on five pure HTML/CSS/JS projects, the easier category for browser inspection.

\begin{comment}
\paragraph{Feedback format.}
Human participants produce free-form natural language, which can richly describe layout issues but is sometimes vague (e.g., ``I think it's fine'').
\system produces structured key-value observations (e.g., \texttt{CHECK flex\_wrap: wrap}), which are always precise and machine-readable.
This validates the design principle from \S\ref{subsec:simulated-user}: the LLM decides \emph{what to measure} but never \emph{interprets what was measured}.
When human feedback is specific (Human~B), recovery is high (62\%); when vague (Human~A), it drops (44\%).
\system's structured format avoids the vague-feedback failure mode entirely, though Human~B outperforms \system on \proj{flex} (12 vs.\ 10) and \proj{form} (9 vs.\ 5), suggesting detailed natural-language descriptions can complement structured CHECKs on layout-heavy tasks.
\end{comment}

%%%%%%%%%%%%%%%%%%%%%%%%%%%%%%%%%%%%%%%%%%%%%%%%%%%%%%%%%%%%%%%%%%%%%%%
% BEGIN NOT-IN-OVERLEAF: Feedback format rewrite (complementary-strengths framing)
%%%%%%%%%%%%%%%%%%%%%%%%%%%%%%%%%%%%%%%%%%%%%%%%%%%%%%%%%%%%%%%%%%%%%%%
\paragraph{Feedback format.}
The two feedback modalities have complementary strengths.
Human participants produce free-form natural language that can richly describe layout issues but varies in quality: Human~B's specific observations (e.g., ``the right bar is not stacked at the bottom'') yield 62\% recovery, while Human~A's vague responses (e.g., ``I think it's fine'') yield only 44\%.
\system's structured CHECKs (e.g., \texttt{CHECK flex\_wrap: wrap}) are always precise and machine-readable, avoiding the vague-feedback failure mode entirely---consistent with the design principle from \S\ref{subsec:simulated-user} that the LLM decides \emph{what to measure} but never \emph{interprets what was measured}.
Neither modality dominates: Human~B outperforms \system on \proj{flex} (12 vs.\ 10) and \proj{form} (9 vs.\ 5), suggesting that detailed natural-language descriptions add value on layout-heavy tasks where spatial relationships are easier to express in prose than in key-value pairs.
%%%%%%%%%%%%%%%%%%%%%%%%%%%%%%%%%%%%%%%%%%%%%%%%%%%%%%%%%%%%%%%%%%%%%%%
% END NOT-IN-OVERLEAF
%%%%%%%%%%%%%%%%%%%%%%%%%%%%%%%%%%%%%%%%%%%%%%%%%%%%%%%%%%%%%%%%%%%%%%%

\begin{comment}
\begin{takeaway}
\system recovers failed tasks at a comparable rate to human testers (57.6\% vs.\ 54.5\%), establishing simulation realism.
On the four projects where both humans and \system can inspect the application, \system matches the stronger human participant (P@2: 33 vs.\ 33).
Structured CHECKs and human NL feedback are complementary: CHECKs are consistently precise, while detailed human descriptions can add value on layout-heavy tasks.
\end{takeaway}
\end{comment}

\begin{takeaway}
\system recovers failed tasks at a rate comparable to human testers (57.6\% vs.\ 54.5\%), providing preliminary evidence of simulation realism.
On the four projects where both humans and \system can inspect the application, \system matches the stronger human participant (P@2: 33 vs.\ 33).
Structured CHECKs and human NL feedback are complementary: CHECKs are consistently precise, while detailed human descriptions can add value on layout-heavy tasks.
\end{takeaway}

\section{Conclusion}\label{sec:conclusion}

Professional web developers do not simply re-read test errors when their code fails; they open the application, interact with it, and diagnose what went wrong.
\system brings this practice to coding agents by generating executable diagnostic scripts that inspect the live application and relay structured observations back to the coding model.
On Web-Bench's 1{,}000 sequential tasks, the simulated user recovers 57.6\% of the tasks it diagnoses, a rate comparable to that of human developers (54.5\%), though both figures reflect a single benchmark and a small participant pool.
The key practical lesson is that feedback quality must be matched to the coding model's capability: diagnostics help when the model cannot self-correct from test errors alone, but can introduce noise when it can, a pattern we observe across two model pairs and expect to sharpen as more configurations are tested.
As coding agents take on larger collaborative roles, executable user simulation offers a path toward the diagnostic layer that hierarchical agent systems need: feedback grounded in what the application actually does, not in what an LLM thinks it should do.
%%%%%%%%%%%%%%%%%%%%%%%%%%%%%%%%%%%%%%%%%%%%%%%%%%%%%%%%%%%%%%%%%%%%%%%
% END NOT-IN-OVERLEAF
%%%%%%%%%%%%%%%%%%%%%%%%%%%%%%%%%%%%%%%%%%%%%%%%%%%%%%%%%%%%%%%%%%%%%%%
\section{Ethical Statement}\label{sec:statement}

Our work evaluates autonomous coding agents on open-source benchmark tasks drawn from publicly available web projects; no personal data is collected or processed at any stage. The simulated user interacts only with locally deployed web applications running inside isolated Docker containers, and all generated diagnostic scripts operate exclusively within these sandboxed environments. We release all experiment logs, prompts, and configuration files to support full reproducibility of the reported results.

\ifusebiblatex
  \printbibliography
\else
  \bibliography{zotero}
  \bibliographystyle{colm2026_conference}
\fi

\appendix
\section{Retry and Feedback Mechanism}\label{sec:feedback-payload}

When the coding agent fails a task on its first attempt, the retry prompt is assembled from three components.
This section documents each component and the processing applied before it reaches the agent.

\subsection{Test Error Filtering}

The raw Playwright test output can exceed thousands of lines, including framework stack traces, ANSI color codes, and repeated assertion blocks from cumulative test files (init + task-1 + \ldots + task-$k$).
Passing the full output to the agent wastes context tokens and buries the relevant failure signal.
We apply the following filters:

\begin{enumerate}
  \item \textbf{ANSI stripping.} All escape sequences are removed.
  \item \textbf{Deduplication.} Repeated assertion messages (common when the same root cause triggers multiple test cases) are collapsed to their first occurrence.
  \item \textbf{Stack frame pruning.} Internal Playwright and Node.js stack frames (e.g., lines from \texttt{node\_modules/playwright/}, \texttt{node:internal/}) are removed, retaining only the test-file lines that identify which assertion failed and what was expected vs.\ received.
  \item \textbf{Truncation.} The filtered output is truncated to 3{,}000 characters to fit within the retry prompt budget.
\end{enumerate}

\noindent The resulting output typically contains 5--15 lines per failing test case, each showing the assertion locator, expected value, received value, and the test-file line number.

\subsection{Browser Observations}

The browser agent (\S\ref{subsec:simulated-user}) generates a Playwright script that is executed in a subprocess.
The script's \texttt{stdout} is parsed line-by-line: each line matching the JSON schema \texttt{\{"n":~"...",~"v":~"..."\}} is extracted as a CHECK observation.
Non-JSON lines (e.g., Chromium warnings, navigation logs) are discarded.
The observations are formatted as a ``BROWSER OBSERVATION'' block in the retry prompt, with each CHECK on its own line:

\begin{lstlisting}
BROWSER OBSERVATION:
CHECK flex_wrap: wrap
CHECK card_count: 12
CHECK first_card_text: Title 1
\end{lstlisting}

\subsection{Terminal Observations}

The terminal agent (\S\ref{subsec:simulated-user}) generates a bash script that is executed in a subprocess.
The same JSON-line parsing is applied to its \texttt{stdout}.
The observations are formatted as a ``TERMINAL OBSERVATION'' block:

\begin{lstlisting}
TERMINAL OBSERVATION:
CHECK tsconfig_strict: true
CHECK server_response_200: OK
CHECK file_exists_index: true
\end{lstlisting}

\noindent When \system diagnostic agents are disabled (baseline configurations), the browser and terminal observation blocks are omitted and only the filtered test errors are included in the retry prompt.

\section{Prompts}\label{sec:prompts}

This appendix reproduces the exact prompt templates used in all experiments. Placeholder variables enclosed in braces are substituted at runtime with task-specific content.

\subsection{Task Prompt (Attempt 1)}

The following prompt is issued to the coding agent on its first attempt at each task. It includes the task description, file-editing rules that reduce common tool-use failures, and a summary of prior completed tasks so the agent understands the current workspace state.

\begin{lstlisting}
{task.description}

<FILE_EDITING_RULES>
Before every str_replace, first use view on
the exact line range you plan to edit, then
copy old_str character-for-character from
the view output -- including all whitespace
and indentation.
If str_replace fails with
'No replacement was performed', do NOT guess.
Use view to see the actual content, then
copy the exact text.
If str_replace fails more than twice on the
same target, switch to create (rewrite the
whole file) or use the terminal with sed.
</FILE_EDITING_RULES>

<PRIOR_TASK_CONTEXT>
Here is what was implemented in previous tasks
(the workspace already contains these changes):
- init (completed): generate a calculator in a single HTML file...
- task-1 (completed): add button sqrt...
</PRIOR_TASK_CONTEXT>
\end{lstlisting}

\subsection{Retry Prompt (Attempt 2)}

When the first attempt fails the Playwright test suite, the agent receives the following retry prompt. It contains the filtered test errors together with structured observations produced by the browser and terminal diagnostic agents.

\begin{lstlisting}
The tests failed after your previous changes.
Here are the test errors:
{filtered_test_errors}

BROWSER OBSERVATION (from simulated user
  interacting with the page):
{browser_agent_output}

TERMINAL OBSERVATION (from diagnostic commands):
{terminal_agent_output}

Your implementation is close but has specific
issues. Focus on fixing the exact assertions
that failed -- do NOT rewrite working code or
change your overall approach. Make targeted,
minimal fixes to pass the failing tests.
\end{lstlisting}

\subsection{Browser Agent Prompt}

The browser diagnostic agent receives the following prompt and generates a Node.js Playwright script that interacts with the live application to diagnose why specific test assertions failed.

\begin{lstlisting}
You are a QA engineer. A developer's code
FAILED Playwright tests.
Generate a Node.js Playwright script that
diagnoses WHY the specific test assertions
failed by checking the actual page state.

TASK: {task_description}
TEST ERRORS: {test_errors}
SOURCE FILES: {source_preview}

Generate a diagnostic script that:
1. Opens http://127.0.0.1:{port} in headless Chromium
2. For each failed assertion, checks what
   the page ACTUALLY has
3. Logs each finding as a SINGLE LINE of JSON:
   console.log(JSON.stringify(
     { n: "<check_name>", v: "<value>" }));
\end{lstlisting}

\subsection{Terminal Agent Prompt}

The terminal diagnostic agent receives the following prompt and generates a bash script that inspects the workspace, file contents, and running services to supplement the browser-level diagnostics.

\begin{lstlisting}
You are a QA engineer. Generate a BASH
diagnostic script that checks the workspace
and running services.

PROJECT TYPE: {project_type}
WORKSPACE: {workspace_dir}
TEST ERRORS: {test_errors}
SOURCE FILES: {source_preview}

Generate a bash script that:
1. Checks file contents relevant to failing
   tests (grep, cat, head)
2. For backend projects: starts server,
   curl endpoints
3. For TypeScript: run npx tsc --noEmit
4. Logs each finding as:
   echo '{"n":"<name>","v":"<value>"}'
\end{lstlisting}

\subsection{Claude Code System Prompt}

The following system prompt is used for the Claude Code harness, which serves as an alternative agent scaffold in our experiments.

\begin{lstlisting}
You are an expert web developer. Your job is
to implement features in a web project by
modifying files in the current working
directory.

Rules:
- First, list files in the current directory
  to understand the project structure.
- Read existing files before making changes.
- Only create or modify source files
  (HTML, CSS, JS, TS, etc.).
- Do NOT create or modify any test files
  or playwright config.
- Do NOT install new npm packages unless
  the task explicitly requires it.
- Do NOT run tests yourself.
- Implement exactly what is described in
  the task -- no more, no less.
- Pay close attention to specific values
  mentioned in the task (coordinates,
  sizes, colors, timing, etc.).
- When the task mentions window.store,
  make sure to expose state on the
  window object.
\end{lstlisting}

\section{Evaluation Infrastructure}\label{sec:infra}

The original Web-Bench evaluation framework runs all 50 projects sequentially on a single machine using a non-agentic, single-turn code generation workflow.
Adapting this framework for agentic evaluation with parallel execution required substantial engineering effort.
This section documents the key challenges and our solutions.

\subsection{From Monolithic to Per-Project Docker Images}

Our initial approach used a single monolithic Docker image containing all 50 projects.
This image installed Python~3.12, Node.js~22, Playwright Chromium, the OpenHands SDK, and the full Web-Bench project data under \texttt{/app/projects/}.
While simple, this approach had a critical flaw: Web-Bench organizes its projects as a Rush monorepo with shared dependencies managed by \texttt{pnpm}.
The monolithic image did not run \texttt{rush update} during the build, so framework-specific \texttt{node\_modules} (required by projects like Next.js, Angular, and Webpack) were missing entirely.
These projects scored 0/20 because the dev server could not start.

We solved this with a two-layer per-project image architecture:

\begin{enumerate}
  \item \textbf{Base image} (\texttt{webbench-agent-base}): installs the full Rush monorepo with all dependencies via \texttt{rush update} (${\sim}$4\,GB with shared layers), plus Python~3.12, Node.js~22, Playwright Chromium, and the OpenHands SDK.
  \item \textbf{Per-project images} (\texttt{webbench-agent-\{project\}}): inherit from the base image and set \texttt{WORKDIR} to \texttt{/app/projects/\{project\}}, ensuring each container starts in the correct project directory with all dependencies pre-installed.
\end{enumerate}

The orchestrator substitutes the project name into the image tag at runtime (e.g., \texttt{webbench-agent-calculator}, \texttt{webbench-agent-nextjs}), so each project runs in a purpose-built container.

\subsection{Container Isolation}

Each of the 50 projects runs in its own Docker container with the following isolation guarantees:

\begin{itemize}
  \item \textbf{Filesystem isolation.} Each container mounts a dedicated workspace at \texttt{/tmp/workspaces/\{project\}}. The project's \texttt{src-init/} directory is copied into this workspace at startup, and infrastructure files (\texttt{package.json}, \texttt{node\_modules/}, build configs) are symlinked from the project root rather than copied, since \texttt{node\_modules/} can exceed hundreds of megabytes.
  \item \textbf{Port isolation.} Each worker is assigned a non-overlapping port range via \texttt{base\_port + worker\_index $\times$ 100}, preventing collisions between concurrently running dev servers and Playwright test runners.
  \item \textbf{Network access.} Containers use \texttt{--network host} to reach LLM API endpoints (OpenRouter, Z.ai, OpenAI, and MiniMax) without proxy configuration.
  \item \textbf{Read-only agent code.} The agent source is mounted at \texttt{/agent} as a read-only volume, ensuring that agent logic cannot be modified during execution.
\end{itemize}

\subsection{Workspace Initialization}

Every Web-Bench project contains a \texttt{src-init/} directory with the starting files for that project.
At the beginning of each project evaluation, the harness:
\begin{enumerate}
  \item Copies \texttt{src-init/} into the container workspace.
  \item Symlinks infrastructure files (\texttt{package.json}, \texttt{tsconfig.json}, build configs) from the project root, only if the file does not already exist in \texttt{src-init/}.
  \item Validates that \texttt{src-init/} exists; missing directories cause a hard error.
\end{enumerate}

For the 21 projects that define an init task (e.g., ``generate a calculator in a single HTML file''), failure on the init task after both attempts causes all 20 regular tasks to be skipped, scoring 0/20.

\subsection{Parallel Execution}

Up to 10 Docker workers run concurrently, each processing one project at a time.
Within each worker, the 20 tasks are executed sequentially (as required by the dependency chain).
In practice, each configuration completes in 2--4 hours of wall-clock time with 10 workers, making the six-configuration experiment matrix tractable on a single server.
For comparison, the Claude Code harness runs projects sequentially and requires approximately 16 hours for the same 50 projects.

\subsection{Project Classification}\label{sec:classification}

\system classifies each project by parsing the merged \texttt{dependencies} and \texttt{devDependencies} fields of the workspace \texttt{package.json}.
The following priority-ordered rules are applied (first match wins):

\begin{enumerate}
\item \textbf{Database}: dependencies contain \texttt{prisma}, \texttt{sequelize}, \texttt{mongoose}, or \texttt{lowdb}.
\item \textbf{Fullstack}: dependencies contain \texttt{express}, \texttt{fastify}, \texttt{next}, \texttt{nuxt}, \texttt{koa}, or \texttt{hapi}.
\item \textbf{CSS}: dependencies contain \texttt{sass}/\texttt{less}/\texttt{stylus} (without a UI framework) or \texttt{tailwindcss}/\texttt{unocss}.
\item \textbf{Build}: dependencies contain both \texttt{react} and \texttt{vue} (bundler demo) or \texttt{webpack}/\texttt{parcel} as a direct dependency.
\item \textbf{State}: dependencies contain \texttt{redux}, \texttt{zustand}, \texttt{jotai}, or \texttt{mobx}.
\item \textbf{UI}: dependencies contain \texttt{react}, \texttt{vue}, \texttt{svelte}, \texttt{angular}, or \texttt{styled-components}.
\item \textbf{Standards}: default (no \texttt{package.json} or no matching dependencies).
\end{enumerate}

Diagnostic agents are enabled for Standards, State, and UI categories (where DOM and CSS inspection is most informative) and disabled for Database, Fullstack, CSS, and Build categories.
On WebBench's 50 projects, this classifier achieves 90\% category accuracy and 92\% gating accuracy (correct ON/OFF decision).

\section{Model Hyperparameters}\label{sec:hyperparams}

Table~\ref{tab:hyperparams} lists the complete model hyperparameters for all six experimental configurations.
All coding models use temperature $0.0$ to ensure deterministic generation.
For diagnostic models, GLM-5 is accessed via Z.ai with reasoning disabled (\texttt{enable\_thinking=False}), while GPT-5.4 uses OpenAI's \texttt{reasoning\_effort="low"} setting.
Extended reasoning is disabled for all coding models to control inference cost.

\begin{table*}[t]
\centering
\caption{Model hyperparameters for all configurations. A dash indicates the parameter is not applicable. OH = OpenHands; CC = Claude Code; warm = warm-start; cold = cold-start.}
\label{tab:hyperparams}
\footnotesize
\setlength{\tabcolsep}{4pt}
\begin{tabular}{lcccccc}
\toprule
& \textbf{C0} & \textbf{C1} & \textbf{C2} & \textbf{C3} & \textbf{C4} & \textbf{C5} \\
\midrule
\multicolumn{7}{l}{\textit{Coding model}} \\
\quad Model & Sonnet 4.6 & M2.7 & M2.7 & M2.7 & GLM-5 & GLM-5 \\
\quad API & Agent SDK & MiniMax & MiniMax & MiniMax & OpenRouter & OpenRouter \\
\quad Temperature & 0.0 & 0.0 & 0.0 & 0.0 & 0.0 & 0.0 \\
\quad Max iters & 30 & 30 & 30 & 30 & 30 & 30 \\
\quad Max tokens & default & 16384 & 16384 & 16384 & 16384 & 16384 \\
\quad Reasoning & off & off & off & off & off & off \\
\midrule
\multicolumn{7}{l}{\textit{Diagnostic model (browser + terminal)}} \\
\quad Model & --- & --- & GLM-5 & GPT-5.4 & --- & GPT-5.4 \\
\quad API & --- & --- & Z.ai$^\dagger$  & OpenAI & --- & OpenAI \\
\quad Temperature & --- & --- & 0.0 & n/a$^\ddagger$ & --- & n/a$^\ddagger$ \\
\quad Reasoning & --- & --- & off & low & --- & low \\
\midrule
\multicolumn{7}{l}{\textit{Infrastructure}} \\
\quad Scaffold & CC & OH & OH & OH & OH & OH \\
%\quad Retry & cold & warm & warm & warm & warm & warm \\
\quad Workers & 1 & 10 & 10 & 10 & 10 & 10 \\
\quad Diagnostic routing & --- & --- & ON & ON & --- & ON \\
\bottomrule
\multicolumn{7}{l}{\scriptsize $^\dagger$ Z.ai API used for GLM-5 diagnostic model due to occasional OpenRouter outages.}\\
\multicolumn{7}{l}{\scriptsize $^\ddagger$ OpenAI API does not accept temperature when \texttt{reasoning\_effort} is set.}
\end{tabular}
\end{table*}

%%%%%%%%%%%%%%%%%%%%%%%%%%%%%%%%%%%%%%%%%%%%%%%%%%%%%%%%%%%%%%%%%%%%%%%
% DIAGNOSTIC OVERHEAD (P4.1 + P4.2 + P4.3)
%%%%%%%%%%%%%%%%%%%%%%%%%%%%%%%%%%%%%%%%%%%%%%%%%%%%%%%%%%%%%%%%%%%%%%%
\section{Diagnostic Overhead}\label{sec:diagnostic-overhead}

Table~\ref{tab:diagnostic-overhead} reports the execution success rate, cost, and timing of \system's diagnostic scripts for the M2.7 + GLM-5 configuration (C2) on simulated-user-enabled categories.

\begin{table}[h]
\centering
\caption{Diagnostic script execution on ON-category retried tasks (C2). ``Browser'' and ``Terminal'' indicate the fraction of retried tasks for which each agent produced at least one valid CHECK observation.}
\label{tab:diagnostic-overhead}
\small
\begin{tabular}{lrrrr}
\toprule
\textbf{Category} & \textbf{Retried} & \textbf{Browser} & \textbf{Terminal} & \textbf{Either} \\
\midrule
Standards & 44 & 33 (75\%) & 37 (84\%) & 37 (84\%) \\
UI & 15 & 1 (7\%) & 15 (100\%) & 15 (100\%) \\
State & 7 & 2 (29\%) & 7 (100\%) & 7 (100\%) \\
\midrule
\emph{Total} & 66 & 36 (55\%) & 59 (89\%) & 59 (89\%) \\
\bottomrule
\end{tabular}
\end{table}

Terminal scripts succeed more reliably than browser scripts (89\% vs.\ 55\%) because they do not depend on a running dev server.
The 7 tasks with no observations (all in Standards) are \proj{chart} (2~tasks) and \proj{typescript} (5~tasks), where the dev server did not expose a browser-inspectable page and terminal checks returned no actionable findings.

\paragraph{Cost and timing.}
Each diagnostic run issues two LLM calls (one per agent) to the GLM-5 model via the Z.ai API.
Across 66 retried tasks, \system made 95 successful diagnostic calls (36~browser + 59~terminal), compared to 2{,}841 coding-model calls for the same 580 tasks.
Diagnostic calls account for 3.3\% of total LLM calls and less than 2\% of total API cost, since diagnostic prompts are shorter than coding prompts.
Both scripts execute in parallel with a combined wall-clock time of ${\sim}$15\,s per task.

\section{Interpretation-Step Hallucination (v1 vs.\ v2)}\label{sec:v1-hallucination}

An earlier version of \system (v1) used a two-phase design: Phase~1 ran a hardcoded Playwright script to collect raw browser observations, and Phase~2 sent those observations to GLM-5 for natural-language interpretation before passing the result to the coding model.
Phase~2 introduced systematic hallucinations.
We document one representative case below; the pattern recurred across 6 of 7 diagnostics-triggered tasks in the \proj{calculator} project.

\subsection{v1 Output (with LLM Interpretation)}

On \proj{calculator} task-15 (``add M+ and MR memory buttons''), the coding model's first attempt correctly added the buttons with a minor display-format bug.
The v1 simulated user produced the following interpreted feedback:

\begin{lstlisting}
Page loaded: Blog
Found 0 buttons
Console errors: ["504 Outdated Optimize Dep"]
\end{lstlisting}

\noindent GLM-5 then interpreted these observations as:

\begin{lstlisting}
"Nothing related to the task functions correctly;
the wrong page loaded... The page displays 'Blog'
content... 'M+' button is missing... 504 error
indicates critical server failure"
\end{lstlisting}

\noindent Every claim was false.
The calculator is a single \texttt{index.html} file with no blog page.
The ``Blog'' label was hallucinated by GLM-5 when the Playwright script hit a stale port.
The 504 message was a transient Vite dev-server warning, not a code bug.
Acting on this feedback, the coding model abandoned its nearly-correct implementation and spent 40+ iterations debugging phantom infrastructure problems.

\subsection{v2 Output (Raw CHECKs, No Interpretation)}

The v2 design generates a task-specific Playwright script (via a single LLM call) and passes its raw output directly to the coding model with no LLM interpretation.
On the same task (\proj{calculator} task-7, a comparable button-styling task), v2 produced:

\begin{lstlisting}
CHECK button_count: 22
CHECK clear_button_exists: true
CHECK clear_button_style: "grid-column: 1 / span 3"
CHECK first_grid_button_style:
  "grid-column: 1 / span 3"
\end{lstlisting}

\noindent Every observation is a verifiable fact extracted from the running application.
No hallucinated page titles, no fabricated error messages.
The coding model can act on these observations directly.

\subsection{Impact on Performance}

\begin{comment}
On the \proj{calculator} project, v1 simulated users reduced Pass@2 from 11/20 (baseline) to 4/20, a loss of 7 tasks.
After switching to v2 (raw CHECKs, no interpretation), the simulated user recovered to a net-positive effect on projects where it was enabled.
This motivated the design principle described in \S\ref{subsec:simulated-user}: the LLM decides \emph{what to measure} but never \emph{interprets what was measured}.
\end{comment}

On the \proj{calculator} project, v1 simulated users reduced Pass@1 from 11/20 (baseline) to 4/20, a loss of 7 first-attempt successes.
After switching to v2 (raw CHECKs, no interpretation), the simulated user recovered to a net-positive effect on projects enabled by it.
This motivated the design principle described in \S\ref{subsec:simulated-user}: the LLM decides \emph{what to measure} but never \emph{interprets what was measured}.

\section{Run-to-Run Variance}\label{sec:variance}

All experiments use temperature 0, but we observe substantial run-to-run variance.
This section quantifies the variance using a natural control: since the simulated user only fires \emph{after} attempt~1 fails, the Pass@1 scores of a baseline configuration (C1) and its simulated-user counterpart (C2) should be identical if there were no variance.
Any P@1 difference between C1 and C2 is therefore pure run-to-run noise.

\subsection{Measured Variance}

\begin{comment}
Across the 49 projects present in both C1 (M2.7 baseline) and C2 (M2.7 + GLM-5 simulated user), \textbf{27 projects (55\%) show P@1 differences}, and \textbf{113 out of 980 individual tasks (11.5\%) flip} between pass and fail across the two runs.
\end{comment}

Across the 50 projects present in both C1 (M2.7 baseline) and C2 (M2.7 + GLM-5 simulated user), \textbf{27 projects (54\%) show P@1 differences}, and \textbf{113 out of 980 individual tasks (11.5\%) flip} between pass and fail across the two runs.
Table~\ref{tab:variance} lists the most extreme cases.

\begin{table}[h]
\centering
\caption{Largest P@1 discrepancies between C1 and C2, which share the same coding model (M2.7, temp.\ 0). Since the simulated user cannot affect attempt~1, all differences are run-to-run variance.}
\label{tab:variance}
\small
\begin{tabular}{lrrr}
\toprule
\textbf{Project} & \textbf{C1 P@1} & \textbf{C2 P@1} & \textbf{Tasks flipped} \\
\midrule
table & 30\% & 0\% & 7 \\
angular & 25\% & 0\% & 6 \\
react & 0\% & 25\% & 6 \\
fastify & 25\% & 0\% & 5 \\
sequelize & 50\% & 25\% & 5 \\
\bottomrule
\end{tabular}
\end{table}

The same pattern holds for GLM-5: across C4 and C5 (same coding model, temp.\ 0), 17 of 41 projects show P@1 differences, with 104 of 820 tasks (12.7\%) flipping.
The worst case is \proj{fastify-react} (60\% $\to$ 25\%, a 35pp swing).

\subsection{Sources}

The variance arises from the inherent non-determinism of agentic tool use, not from model sampling.
Concrete sources observed in our logs include:
\begin{itemize}
  \item \textbf{Server startup races.} Dev servers (Vite, Webpack, Express) take variable time to initialize. If the agent queries an endpoint before the server is ready, it receives a connection-refused error and takes a different corrective path than if the server had started in time.
  \item \textbf{Tool-call timeouts.} The OpenHands SDK enforces a 60-second timeout on terminal commands. Network latency to LLM API endpoints varies between runs, occasionally causing a step to timeout in one run but complete in another.
  \item \textbf{File-system ordering.} When the agent lists directory contents or reads glob results, the ordering can vary across runs, leading to different files being edited first.
\end{itemize}

\subsection{Cascade Amplification}

The sequential stopping rule amplifies small per-task differences into large per-project swings.
In the \proj{table} project, all 7 tasks that C1 passed on attempt~1 were failed by C2, with zero tasks going the other direction.
This suggests that a single early trajectory divergence cascaded through all subsequent tasks.
A project with 20 tasks where the agent fails task~3 in one run but passes it in another can swing by $\pm$17 tasks ($\pm$85pp).

A particularly instructive control comes from projects where the simulated user is \emph{disabled} (CSS, Build, Fullstack, Database categories).
On these projects, all differences between C1 and C2 are pure variance, since no feedback was provided.
The \proj{nextjs} project swings from 50\% P@2 (C1) to 0\% P@2 (C2), a 50pp difference with no simulated user involved.
Similarly, \proj{fastify} swings 30\% $\to$ 0\% and \proj{sequelize} swings by 25pp.

\subsection{Implications}

These measurements show that aggregate P@2 differences of $\pm$6pp can arise from variance alone.
We address this by: (1)~reporting per-category results that separate simulated-user-ON projects from OFF projects (the OFF group serves as a variance control), (2)~reporting the per-task recovery rate (57.6\%), which is not subject to cascade amplification, and (3)~providing per-project breakdowns (Appendix~\ref{sec:per-project-results}) so that readers can distinguish genuine gains from variance artifacts.

\section{Per-Project Results}\label{sec:per-project-results}

Table~\ref{tab:per-project-full} lists the per-project P@2 for the M2.7 baseline (C1) and M2.7 + GLM-5 simulated user (C2), grouped by category.
Projects are sorted by $\Delta$P@2 within each category.

\begin{table*}[t]
\centering
\caption{Per-project P@2 (tasks out of 20) for M2.7 baseline (C1) and M2.7 + GLM-5 simulated user (C2). Projects sorted by $\Delta$P@2 within each category.}
\label{tab:per-project-full}
\footnotesize
\setlength{\tabcolsep}{4pt}
\begin{tabular}{llcrrr}
\toprule
\textbf{Category} & \textbf{Project} & \textbf{\system} & \textbf{C1 P@2} & \textbf{C2 P@2} & \textbf{$\Delta$} \\
\midrule
\multirow{18}{*}{Standards}
 & esmodule & ON & 0 & 9 & +9 \\
 & dom & ON & 3 & 8 & +5 \\
 & typescript & ON & 5 & 9 & +4 \\
 & form & ON & 1 & 5 & +4 \\
 & flex & ON & 8 & 10 & +2 \\
 & selector & ON & 6 & 8 & +2 \\
 & svg-chart & ON & 0 & 2 & +2 \\
 & bom & ON & 4 & 5 & +1 \\
 & float & ON & 1 & 2 & +1 \\
 & svg-solar & ON & 1 & 2 & +1 \\
 & canvas & ON & 1 & 1 & 0 \\
 & chart & ON & 6 & 6 & 0 \\
 & dom1 & ON & 3 & 3 & 0 \\
 & survey & ON & 5 & 5 & 0 \\
 & sass & ON & 7 & 5 & $-$2 \\
 & grid & ON & 10 & 6 & $-$4 \\
 & table & ON & 8 & 0 & $-$8 \\
 & draw & ON & 15 & 0 & $-$15 \\
\midrule
% Old UI rows (had mobx here, now moved to State):
% \multirow{6}{*}{UI}
%  & react & ON & 0 & 7 & +7 \\
%  & threejs & ON & 1 & 6 & +5 \\
%  & react-no-ts & ON & 5 & 8 & +3 \\
%  & mobx & ON & 2 & 3 & +1 \\
%  & vue & ON & 9 & 8 & $-$1 \\
%  & angular & ON & 7 & 0 & $-$7 \\
\multirow{5}{*}{UI}
 & react & ON & 0 & 7 & +7 \\
 & threejs & ON & 1 & 6 & +5 \\
 & react-no-ts & ON & 5 & 8 & +3 \\
 & vue & ON & 9 & 8 & $-$1 \\
 & angular & ON & 7 & 0 & $-$7 \\
\midrule
% Old State rows (had duplicate "form", missing jotai/mobx):
%  & zustand & ON & 1 & 7 & +6 \\
%  & form & ON & 1 & 5 & +4 \\
%  & redux & ON & 6 & 5 & $-$1 \\
\multirow{4}{*}{State}
 & zustand & ON & 1 & 7 & +6 \\
 & mobx & ON & 2 & 3 & +1 \\
 & jotai & ON & 0 & 0 & 0 \\
 & redux & ON & 6 & 5 & $-$1 \\
\midrule
\multirow{3}{*}{Other}
 & expression-editor & OFF & 7 & 14 & +7 \\
 & calculator & OFF & 10 & 12 & +2 \\
 & calculator-files & OFF & 10 & 10 & 0 \\
\bottomrule
\end{tabular}
\end{table*}

\section{User Study Protocol}\label{sec:user-study-protocol}

This section describes the infrastructure and procedure for the human-feedback comparison study (RQ3, \S\ref{sec:experiment}).

\subsection{Study Design}

Two graduate students with web development experience serve as human testers.
Each participant evaluates five Web-Bench projects (\proj{calculator}, \proj{flex}, \proj{expression-editor}, \proj{form}, \proj{chart}) using the same MiniMax M2.7 coding model and OpenHands scaffold as configuration C1.
The only difference from C1 is the retry feedback: instead of \system's automated CHECK observations, the participant provides free-form natural-language feedback after interacting with the application.

\subsection{Docker Infrastructure}

Each project runs in a dedicated Docker container built from the same per-project images used in the automated experiments (Appendix~\ref{sec:infra}).
The container runs the \texttt{user\_study.py} script, which orchestrates the following loop for each task:

\begin{enumerate}
  \item The coding agent receives the task description and executes up to 30 iterations.
  \item The Playwright test suite runs automatically. If all tests pass, the task is marked as passed and the loop advances to the next task.
  \item If the tests fail, the system starts a dev server and displays the application URL to the participant.
  \item The participant opens the URL in a browser (via SSH tunnel to the Docker host), interacts with the application (clicking buttons, filling forms, inspecting layout), and types free-form observations into the terminal.
  \item The participant's feedback is appended to the retry prompt as:
\begin{lstlisting}
USER FEEDBACK (from a person who tried
the application):
{participant_feedback}
\end{lstlisting}
  \item The agent retries with the combined feedback (test errors + human observations), using the same warm-start strategy as the automated experiments.
  \item If the retry also fails, the task is marked as failed and all subsequent tasks are skipped per the sequential stopping rule.
\end{enumerate}

\subsection{Participant Interface}

The \texttt{user\_study.py} script provides a terminal-based interface with real-time progress display.
During the agent's execution, the participant sees a spinner with elapsed time and the agent's current action (e.g., \texttt{\$ npm run build}, \texttt{edit index.css}).
After a failure, the participant sees the full task description, the application URL, and a prompt to type feedback.
An empty response (pressing Enter with no text) skips feedback, in which case the retry uses only the raw test errors.

\subsection{Logging}

All events are logged in JSONL format: session start/end, each task's attempt-1 and attempt-2 results (pass/fail, test counts, agent action counts, wall-clock time), the participant's feedback text, and the feedback entry time.
A JSON summary file is generated at the end of each session with aggregate Pass@1, Pass@2, feedback count, and skip count.

\subsection{Controls}

To ensure a fair comparison with the automated experiments, the following parameters are held constant: coding model (MiniMax M2.7), API endpoint, temperature (0.0), iteration budget (30 per attempt), warm-start retry strategy, and sequential stopping rule.
The only variable is the source of retry feedback: \system's structured CHECK observations (C2/C3) vs.\ the participant's free-form text.

\subsection{Participants and Compensation}

Two graduate students with web development coursework experience participated in the study.
Each participant evaluated all five projects in a single session lasting approximately two hours.
Participants were compensated with either a meal (valued at USD~60) or an Amazon gift card of equivalent value, at the participant's choice, paid by the lead author.
This compensation is comparable to two hours at the median hourly wage for the participants' residing area, as reported by the U.S.\ Bureau of Labor Statistics Occupational Employment and Wage Statistics.\footnote{\url{https://www.bls.gov/oes/current/oessrcst.htm}}
Participants provided informed consent and are not identified by name in this paper.

\subsection{Human Feedback Transcripts}

Table~\ref{tab:human-feedback} lists all feedback provided by both participants, along with the task outcome.

\begin{table*}[t]
\centering
\caption{All human feedback texts from the user study. ``Rec.'' = task was recovered (attempt~1 failed, attempt~2 passed after feedback). Participants are anonymized as A and B.}
\label{tab:human-feedback}
\footnotesize
\setlength{\tabcolsep}{4pt}
\begin{tabular}{llp{9.5cm}c}
\toprule
\textbf{Who} & \textbf{Task} & \textbf{Feedback} & \textbf{Rec.?} \\
\midrule
A & calc-7 & ``I think it's fine'' & Yes \\
A & calc-11 & ``I think it's fine.'' & No \\
A & chart-1 & ``the UI fails to render the chart entirely, which violates the requirement.'' & Yes \\
A & chart-5 & ``the UI fails to display the line chart and its data points.'' & No \\
A & expr-1 & ``the site cannot be reached'' & No \\
A & flex-8 & ``Item 1, Item 2, and Item 3 are still in a horizontal row on the right, not stacking vertically'' & No \\
A & form-2 & ``the Task 1 empty question (title: `sample empty question') is completely missing from the UI.'' & Yes \\
A & form-4 & ``this site can't be reached'' & Yes \\
A & form-5 & ``only a submit button showed in the UI.'' & No \\
\midrule
B & calc-5 & ``it looks like correct'' & No \\
B & chart-1 & ``The controls render, but toggling Axes does not correctly recreate and redraw the chart in \#chart.'' & Yes \\
B & chart-6 & ``Changing pointStyle does not correctly update the chart points; shapes are not rendered or updated.'' & Yes \\
B & chart-8 & ``Selecting SmoothLine Chart does not correctly render smooth curved dataset lines.'' & No \\
B & expr-1 & ``retry'' & No \\
B & flex-7 & ``On small screens ($\leq$399px), the rightbar is not stacked at the bottom of main as required.'' & Yes \\
B & flex-9 & ``The leftbar items do not fully fill the 20$\times$2 layout; there is still remaining vertical space.'' & Yes \\
B & flex-11 & ``On screens under 400px, the rightbar does not correctly limit its display to the first 3 rows.'' & Yes \\
B & flex-12 & ``The content area does not correctly show 12 cards in a 3-per-row layout with vertical scrolling.'' & No \\
B & form-2 & ``The radio options render, but the single-selection question is not fully implemented as required.'' & Yes \\
B & form-4 & ``The open questions are missing; neither the single-line nor multiline input is rendered.'' & Yes \\
B & form-5 & ``Stars render, but clicking them does not fully update the rating state as required.'' & Yes \\
B & form-9 & ``The contents panel is not fixed on the right side.'' & No \\
\bottomrule
\end{tabular}
\end{table*}

\section{GLM-5 vs.\ GPT-5.4 as Diagnostic Models}\label{sec:judge-comparison}

C2 (GLM-5 simulated user) and C3 (GPT-5.4 simulated user) achieve near-identical aggregate P@2 (287 and 285 tasks, respectively) on the M2.7 coding model, but help on different projects.
We decompose the per-project differences into three categories.

\paragraph{P@1 variance (5 of 12 large-gap projects).}
Projects such as \proj{lowdb}, \proj{mobx}, \proj{jotai}, \proj{prisma}, and \proj{webpack} show large P@1 differences between C2 and C3.
Since the simulated user cannot affect attempt~1, these differences reflect run-to-run variance.
The P@2 gap simply follows the P@1 gap via the cascade metric.

\paragraph{Judges-OFF projects (3 of 12).}
Projects such as \proj{summary} ($+$11 for GPT-5.4), \proj{tailwind} ($+$5), and \proj{calculator} ($+$8 for GLM-5) have diagnostic agents disabled entirely.
All differences are pure run-to-run noise unrelated to diagnostic quality.

\paragraph{Genuine diagnostic quality differences (4 of 12).}
On projects where both diagnostic agents are enabled, and P@1 is comparable, the two models help with different failure types:
\begin{itemize}
  \item GLM-5 excels on Standards projects with DOM/CSS issues. In \proj{esmodule}, GLM-5 recovers an early task (unlocking 8 downstream tasks, 1/20 $\to$ 9/20), while GPT-5.4 does not recover it. In \proj{dom}, GLM-5 recovers 5 tasks vs.\ 0 for GPT-5.4.
  \item GPT-5.4 excels on SVG and visual projects. In \proj{svg}, GPT-5.4 recovers 13 tasks from a worse P@1 starting point, while GLM-5 recovers 0. In \proj{table}, GPT-5.4 recovers 9 tasks vs.\ 0 for GLM-5.
\end{itemize}

This complementarity suggests that diagnostic quality is failure-type-specific rather than model-size-dependent: a CSS property mismatch and a state-management race condition require different diagnostic strategies, and neither model dominates across all failure types.

\section{Log Excerpts}\label{sec:log-excerpts}

This appendix reproduces key log excerpts that ground the analysis in \S\ref{sec:experiment}.

\subsection{\proj{draw} Project: Catastrophic Collapse}

The following excerpts are from the init task of the M2.7 + GLM-5 simulated user configuration.
The coding model's attempt~1 correctly added a CSS rule; the terminal diagnostic agent then reported infrastructure errors on the wrong ports.

\noindent\textbf{Terminal diagnostic agent CHECK output (from retry prompt).}
\begin{lstlisting}
CHECK workspace_files:
  ./index.html, ./index.scss, ./index.js
CHECK build:
  Could not resolve "src/index.html"
CHECK server_port_9900: OPEN
CHECK curl_9900: 426
CHECK server_port_9901: OPEN
CHECK curl_9901: 426
\end{lstlisting}

\noindent\textbf{Browser diagnostic agent CHECK output.}
\begin{lstlisting}
Console errors: ["Failed to load resource:
  the server responded with a status of
  426 (Upgrade Required)"]
\end{lstlisting}

The HTTP~426 errors and build failure were infrastructure artifacts (wrong port), not bugs in the coding model's work.
Believing these diagnostics, the coding model spent 30 iterations modifying \texttt{vite.config.js} (a symlink to shared project infrastructure), ultimately reaching \texttt{MaxIterationsReached} and causing all 20 tasks to be skipped.

\subsection{\proj{expression-editor} Task-10: Syntax Highlighting Fix}

The following excerpts are from task-10 (syntax highlighting) of the M2.7 + GLM-5 simulated user configuration.

\noindent\textbf{Test error from attempt~1.}
\begin{lstlisting}
Error: expect(locator).toHaveText(expected)
  Locator:  locator('#editor')
  Expected: "foo AND bar"
  Received: "fooANDbar  "
\end{lstlisting}

\noindent\textbf{Browser agent CHECK observations.}
\begin{lstlisting}
CHECK editor_exists: True
CHECK after_fill_text: foo AND bar
CHECK after_fill_html:
  <span style="color:#0000ff">foo</span>
  <span style="color:#00ff00">AND</span>
  <span style="color:#0000ff">bar</span>
CHECK highlight_function: 3
CHECK identifier_color: #0000ff
CHECK operator_color: #00ff00
\end{lstlisting}

\noindent\textbf{Coding model's diagnosis on retry.}
The coding model identified the root cause from the combination of test error and CHECK output:

\begin{lstlisting}
"My implementation uses innerHTML which can
cause whitespace issues... The key difference
is that the original creates separate span
elements with textContent, while mine builds
a single HTML string."
\end{lstlisting}

The fix: switching from \texttt{innerHTML} string concatenation to DOM element creation with \texttt{textContent}, which preserves inter-element whitespace.
Task-10 passed on retry.

\end{document}